\documentclass[fleqn,twoside]{article}
\usepackage{espcrc2}

\usepackage{graphicx}

\newcommand{\mres}{m_{\rm res}}
\newcommand{\GeV}{\rm GeV}

\title{Chiral properties of domain wall fermions  with improved gauge actions}

\author{Konstantinos Orginos
\address{RIKEN-BNL Research Center,Bldg 510a, Upton, NY 11973-5000}
[RBC Collaboration]
\thanks{We thank RIKEN, Brookhaven National Laboratory and the U.S.\ Department
of Energy for providing the facilities essential for the completion of
this work.}}
 
\begin{document}

\begin{abstract}
 We study the chiral properties of quenched domain wall fermions with
several gauge actions. We demonstrate that the nearly translationally
invariant modes in the fifth dimension that dominate the residual mass
for Wilson gauge action can be substantially suppressed using improved
gauge actions. In particular, we study the Symanzik action, the
Iwasaki action, the DBW2 action and compare them to the Wilson action.
\vspace{1pc}
\end{abstract}

\maketitle

\section{INTRODUCTION}

 Domain wall fermions provide a very elegant solution to the problem
of chiral fermions on the lattice; in the limit of infinite length
$L_s$ of the fifth dimension chiral symmetry becomes exact even when
the lattice spacing does not vanish.  Since practical
implementations can only deal with finite $L_s$ some residual chiral
symmetry breaking is induced. This breaking goes  to
zero as $L_s$ goes to infinity.  It is therefore important to chose
the parameters of the simulation so that an acceptably small chiral
symmetry breaking is achieved at a reasonably small $L_s$.

 Recently, RBC~\cite{Wu:1999cd} and
CP-PACS~\cite{AliKhan:1999zn,AliKhan:2000iv} reported that by using
the Iwasaki gauge action, rather than the Wilson action, a significant
reduction of the residual chiral symmetry breaking is achieved for the
same $L_s$ and lattice spacing $a$.  It is instructive therefore to
study the effects of the gauge action in a systematic way.

 We report here our results from our studies of the residual chiral
symmetry breaking for the 1-loop Symanzik, 
Iwasaki~\cite{Iwasaki:1983ii}, and   DBW2 actions.
The DBW2 action  was introduced
in~\cite{Takaishi:1996xj} and it was shown by QCD-TARO
in~\cite{deForcrand:1999bi} that it is a good approximation of the RG
flow on the two dimensional plane of the plaquette and rectangle
couplings.

All the above gauge actions can in general be written as
\begin{eqnarray}
 S_G &=& \frac{\beta}{3}  
   \left(  c_0 \sum_{x;\mu<\nu} P_{\mu\nu}
        + c_1 \sum_{x;\mu\neq\nu} R_{\mu\nu} + \right. \nonumber\\
      & &  \left.+ c_2 \sum_{x;\mu<\nu<\sigma}  C_{\mu\nu\sigma}
  \right)
 \label{eq:GaugeAct}
\end{eqnarray}
where $P_{\mu\nu}$ is the standard plaquette in the $\mu,\nu$ plane, and
 $R_{\mu\nu}$ and $C_{\mu\nu\sigma}$ denote the real part of the trace 
of the ordered product of SU(3) link matrices along $1\times 2$ rectangles and $1\times 1 \times 1$ paths, respectively. 
For the  1-loop Symanzik action the coefficients $c_0$, $c_1$, and $c_2$
are computed in tadpole improved one loop perturbation theory~\cite{Alford:1995hw}.
While $c_2=0$ and $c_0=1-8c_1$ for both Iwasaki and DBW2,
$c_1=-0.331$ and $-1.4067$ respectively.

As a measure of the chiral symmetry breaking we use the so called
residual mass. The residual mass is defined as
\begin{equation}
\mres = \left.\frac{\langle J^5_q(0)J^5(t)\rangle}
               {\langle J^5(0)J^5(t)\rangle}\right|_{t \ge t_{min}},
\label{eq:Mres}
\end{equation}
where $J^5_q$ is the mid-point chiral symmetry breaking term which
appears in the axial Ward identity of domain wall fermions, and
$t_{min}$ is sufficiently large to avoid short-distance lattice
artifacts.

In addition, for each configuration we have defined the ratio
\begin{equation}
R(t) =  \frac{ J^5_q(0)J^5(t)}
             { J^5(0)J^5(t) },
\label{eq:cnfMres}
\end{equation}
which is very useful in monitoring the chiral symmetry breaking
on each configuration.

\section{MEASUREMENTS - RESULTS}

We measured the light hadron spectrum, the residual mass, and the
ratio $R(t)$ of Eq.~\ref{eq:cnfMres} on $16^3\times32$ lattices. We used
50 Symanzik and Iwasaki lattices, and 90 DBW2 lattices. For
all the actions tested  the lattice spacing
was tuned to $a^{-1}\sim 2\GeV$ using the $\rho$ mass at the chiral
limit to set the scale. This matches the Wilson gauge action at $\beta
= 6.0$.  The corresponding $\beta$ values are given in
Table~\ref{table:sim}.  We have optimized the value of the domain wall
height $M_5$. For all actions except the DBW2, the optimum value of
$M_5$ was found to be 1.8. For the DBW2 action the optimum $M_5$ is
1.7, which  is what we used for the results reported below.

\begin{table}[t]
\caption{The simulation parameters for all the actions}
\label{table:sim}
\newcommand{\m}{\hphantom{$-$}}
\newcommand{\cc}[1]{\multicolumn{1}{c}{#1}}
\begin{tabular}{@{}lllll}
\hline
Action        & \cc{$\beta$} & \cc{$M_\rho$} & \cc{$M_5$} & \cc{$L_s$} \\
\hline
Wilson~\mbox{\cite{Blum:2000kn}}        & \m6.00 & 0.404(8)  & 1.8 & 12-24 \\
Symanzik      & \m8.40 & 0.411(14) & 1.8 & 16 \\
Iwasaki       & \m2.60 & 0.415(13) & 1.8 & 16 \\
DBW2          & \m1.04 & 0.399(11) & 1.7 & 8-16  \\
\hline
\end{tabular}\\[2pt]
\end{table}

\begin{figure}[t]
\vspace{9pt}
\includegraphics[width=70mm]{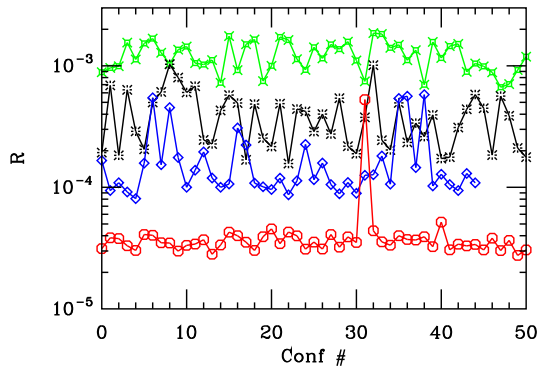}
\caption{
$R(t)$ for bare quark mass 0.02
summed over t from  4 to 16. 
The fancy squares correspond to Wilson action, the bursts to
Symanzik, the diamonds to  Iwasaki and the octagons to  DBW2.}
\label{fig:Spikes}
\end{figure}

In Figure~\ref{fig:Spikes} we plot the the ratio of
eq.~\ref{eq:cnfMres} summed over time slices 4 to 16. As we can see,
the Wilson and Symanzik action data (squares and stars) show very
large fluctuations. But the Iwasaki and DBW2 show a baseline of small
fluctuations interrupted by spikes. Presumably the large fluctuations
observed for the Wilson and Symanzik action are due to a very high
frequency of spikes. The important feature of Figure~\ref{fig:Spikes}
is that the number of spikes is drastically reduced for the DBW2
action.  It is expected that these spikes are spatially localized and
occur on configurations that have eigenvalues of the transfer matrix
in the 5th dimension very close to 1.  We have confirmed these
expectations by measuring both the spacial structure of the spikes and
studying the spectral flow of the 4D Wilson Dirac operator on these
configurations (see talk by T. Izubuchi at this conference).

\begin{figure}[t]
\vspace{9pt}
\includegraphics[width=70mm]{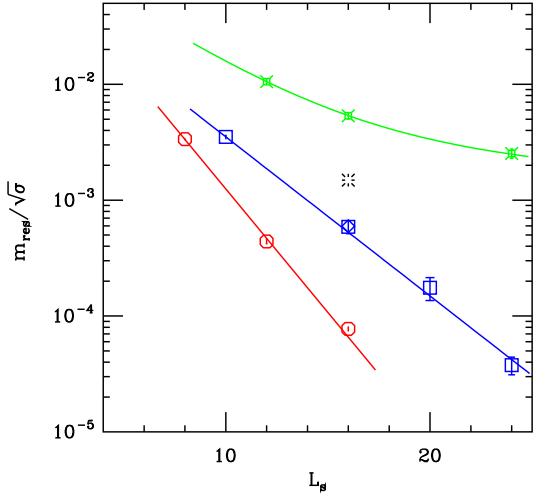}
\caption{$L_s$ dependence of $\mres$ at $a^{-1}\sim2\rm{GeV}$.  The
octagons correspond to the DBW2 action, the squares
(CP-PACS~\cite{AliKhan:2000iv}) and diamond (RBC) to Iwasaki, 
the burst to Symanzik, and the fancy squares to
Wilson.}
\label{fig:Mres_vs_ls}
\end{figure}

In Fig.~\ref{fig:Mres_vs_ls} we show the $L_s$ dependence of the
residual mass. All data are for $a^{-1} = 2\rm{GeV}$ and bare quark
mass $m_q = 0.02$.  The value of $\frac{M_{\pi}}{M_{\rho}}$ is within
$10\%$ of $.55$. Since we are interested in differences by factors of
ten, and since we know that the $m_{res}$ depends mildly on the quark
mass~\cite{Blum:2000kn,AliKhan:2000iv}, the data can be directly
compared as is. For completeness we also present the CP-PACS residual
mass results~\cite{AliKhan:2000iv}.  We have confirmed their results
for $L_s=16$ (diamond in Fig.~\ref{fig:Mres_vs_ls}).  For the Symanzik
action we ran at $L_s=16$ only; the residual mass turns out to be
smaller by a factor of three than the Wilson residual mass at the same
$L_s$, and it is larger by a factor of two than that of the Iwasaki
action.  The DBW2 action not only gives the smallest residual mass,
but also has the steepest decrease as a function of $L_s$. Since we
have only three points, we cannot claim that we know the asymptotic
behavior; it is interesting to note however, that a simple fit gives
us $m_{res}(s) \sim q^s$ with $q \sim 0.6$. Shamir's one loop
perturbative~\cite{Shamir:2000cf} result is $q=0.5$.

For the DBW2 action, in addition to the residual mass tests we have
also measured the light hadron spectrum and pseudoscalar decay
constants at both $2\GeV$ and $1.3\GeV$. The residual mass at
$1.3\GeV$ is about a factor of two smaller than the Wilson residual
mass at $2\GeV$ ($L_s=16$). This allows simulations at coarser lattice
spacing and at moderate values of $L_s$ with minimal chiral symmetry
breaking. Our spectrum results show that the already very good scaling
of domain wall fermions with Wilson gauge action is preserved with the
DBW2 action.  For a detailed presentation of our spectrum data see the
talk by Y. Aoki at this conference.

 It has been argued~\cite{Narayanan:1998yv} that the enhancement of
chiral symmetry breaking is caused by configurations which also
represent changing topology ({\it i.e.} small instantons and/or
dislocations). Our data (Fig.~\ref{fig:Spikes}) indicate that both the
Iwasaki and the DBW2 actions suppress these configurations.  In fact,
Iwasaki~\cite{Iwasaki:1983ii} has argued that his choice of the $c_1$
coefficient is on the boundary separating the space of actions which
have stable instantons and those which have unstable instantons.
Consequently, when these improved actions are used, the question
whether the topology changes arises. We have measured the topological
charge on all the actions we tested, and we have concluded that the
topology changes with a significantly lower rate for the DBW2
action. The problem gets more severe as we get closer to the continuum
limit. The Iwasaki action also shows signs of slow topology changing.
We believe that this slowdown in the topology change can be overcome by
using over-relaxed heatbath algorithm and by simply running for a long
time before saving configurations.  Given the considerable cost of
measuring fermionic observables with DWF we can afford to do so.

\section{CONCLUSIONS}
Using the DBW2 action, we have essentially eliminated the problem of
residual chiral symmetry breaking for  domain wall fermions at
$2\rm{GeV}$ with $L_s=16$.  The DBW2 residual mass is about an order of
magnitude smaller than the Iwasaki $\mres$, and two orders of
magnitude smaller than the Wilson $\mres$.  Assuming a $440\rm{MeV}$
string tension, the bare DBW2 residual mass is about $30\rm{KeV}$. The
observed topology change slowdown may be handled by using longer
running between measurements. Finally, our results suggest that
the DBW2 action should also be useful for  overlap fermions, since it
probably makes the approximation to the sign function converge faster
by eliminating the small eigenvalues of the Hermitian Wilson Dirac
operator.

\end{document}